\documentclass[preprint,showpacs,aps]{revtex4}
\begin{document}
\draft
\newcommand{\be}{{\,\bf e}}
\newcommand{\bp}{{\bf p}}
\newcommand{\bq}{{\bf q}}
\newcommand{\br}{{\bf r}}
\newcommand{\bv}{{\bf v}}
\newcommand{\bE}{{\bf E}}
\newcommand{\bF}{{\bf F}}
\newcommand{\bH}{{\bf H}}
\newcommand{\bJ}{{\bf J}}
\newcommand{\bM}{{\bf M}}
\newcommand{\dfrac}{\displaystyle\frac}
\newcommand{\tJ}{{\hat{\bf J}_{\rm M}}}

\title{Current induced magnetization switching in magnetic tunnel
junctions}
\author{Z. F. Lin, S. T. Chui and L. B. Hu}
\affiliation{ Bartol Research Institute,
University of Delaware, Newark, DE 19716}

\date{\today}

\begin{abstract}
A new mechanism different from the spin accumulation picture is proposed for
the current induced magnetization
switching in magnetic tunnel junctions by
taking into account
the effect of the electron electron interaction.
We found in tunnel structures the possibility of
an enhanced spin switching effect
that, when normalized with respect to the current,
is much bigger than that in multilayers.
Some recent experimental results show evidence for the present picture.
\end{abstract}

\pacs{PACS numbers:73.40.-c,71.70.Ej,75.25.+z}
\maketitle

Spin polarized transport such as the giant magnetoresistance (GMR)
and the tunnelling magnetoresistance (TMR) has recently received
much interest \cite{Book,Berger,Slon1,Cornell,Tsoi,1,Kohl}.
This is partly motivated by technological
applications such as magnetic sensors, hard disk read heads,
magnetic non-volatile random access memories (MRAM) and the possibility of
integrating magnetic elements into semiconductor microelectronics
circuits. The crucial physics that goes into the design of these
structures is that of the transport of electrons over distances
short enough that the spin memory has not been lost. One of the
effects that is manifested is that of spin induced switching of
magnetization.

When a polarized electric current goes from one magnetic element
to the next, the magnetization carried by it when it leaves the first element
will exert a torque and rotate the
magnetization of the second element. This is the spin accumulation
picture for magnetization switching.
Following earlier theoretical suggestions \cite{Berger,Slon1}
a number of experiments \cite{Cornell,Tsoi} have recently
demonstrated this effect under {\bf high current densities}.
If the current
required can be reduced this rotation may
provide a way to {\bf write} a bit in a memory device.
In this Letter, a different mechanism is proposed for magnetization
switching in magnetic tunnel junction
which is expected to be realizable
under much lower current densities.

Intuitively, for tunnel junctions, the resistance is higher, the current and
the magnetization carried by it is reduced. According to the spin
accumulation picture the magnetization switching effect will be reduced.
However, there
are additional physics that needs to be included, which leads to a new
mechanism of the magnetization switching effect.
The tunnel junction can be thought of as a capacitor with
charges localized at the metal-insulator interfaces within a
{\bf screening} length $\lambda$. To consider the capacitance
correctly it is essential to include the effect of
the electron-electron ($e$-$e$) interaction in the nonequilibrium
transport of the system.
We found that very interesting physics happens after taking into account
the $e$-$e$ interaction. There are two
very different length scales in the problem: The bare spin
diffusion length $l_{sf}$, of the order of 100 \AA \ or more, is
much larger than the bare screening length $\lambda_0$ which is of
the order of an \AA.
Under steady state {\bf non-equilibrium conditions}, the external
voltage induced a non-uniformed charge dipole layer at the
metal-insulator interfaces.  These charges are sums of two terms.
In addition to one decaying with a length scale of the order of
the screening length that one encounters in the equilibrium
situation, there is another one decaying with a length scale of
the order of the spin diffusion length. The magnitude of the
density of the latter is smaller than that of the first but they
contribute equally to the screening electric field due to the
long range nature of the Coulomb potential. {\bf In addition,
there is a magnetization dipole layer also induced at the
interfaces}. The magnitude of this magnetization dipole layer is
much larger than that of the charge dipole layer by a factor of
$l_{sf}/\lambda_0$. This magnetization dipole layer is zero under
equilibrium conditions and is nonzero only under steady state
conditions. It generates an effective
switching field. In particular, the magnitude of the induced
magnetization is controlled by the {\bf voltage, not the current},
implying a possibility of achieving an enhanced switching effect
in tunnel structures
that, when normalized with respect to the current, is
much bigger than that caused by the spin accumulation picture.
We thus expect this to have a strong effect in tunnel junctions
where the voltage involved is higher while the current density remains low.
There is some experimental evidence \cite{Kohl}
for the present mechanism, although
definitive studies remain to be carried out.
We now describe our results in detail.

The system we have in mind is a ferromagnetic tunnel junction
where the two interfaces between the ferromagnet-insulator-ferromagnet
sandwich structure are
assumed to be at $z=\pm d/2$.
We assume the $z$ axis to be perpendicular to the
faces of the tunnel junction.
The initial magnetization are assumed
to be in the $x$-$y$ plane with an orientation given by
$\bp_0^R=\cos\Phi \be_x + \sin\Phi\be_y$ for the ferromagnet on the right
hand side and $\bp_0^L=\be_x$ for the ferromagnet
on the left hand side of the sandwitch structure. Our strategy is to assume
magnetization uniform in the $x$-$y$ plane on both the left and the right,
consider a current going through
this junction and calculate the magnetization change $\delta \bM$
under steady state nonequilibrium conditions. The effective switching
field is then estimated by as $H_s=\delta \bM/\chi$ where $\chi$
is the magnetic susceptibility.
Because the work functions of the ferromagnets
on opposite sides of the junction may not be equal, at zero external bias there
will be a charge dipole layer formed at the interfaces. What we
are calculating here are the changes from the zero bias situation.
The experimental structures usually
possess edge domains where the switching starts. The
magnetization is thus not completely uniform in the $x$-$y$ plane.
To bring out the
essential physics, we shall not consider this complication
in the present paper but we hope to come
back to this in the future.

Our starting point is the equation of motion of the charge and the
magnetization. For the charge, it is just the equation of charge
current conservation
\begin{equation}
\nabla\cdot\bJ_e=-\frac{\partial \delta\! n}{\partial t}
\end{equation}
where $\bJ_e$ is the total current.
For the magnetization $\bM$, the equation takes
the form of the classical Landau-Lifshitz equation
\begin{equation}
\frac{\partial \bM }{\partial t} - \gamma \bM\times\bH +
 \nabla\cdot\tJ = -\frac{\delta\!\bM}{\tau}
\end{equation}
where $\gamma$ is the
gyromagnetic ratio,
$\tJ$ is the spin current
(tensor), and $\bH$ is the effective field
describing the precession of the magnetic moments given by
$\bH=\bH_{{\rm e}}+\bH_{{\rm an}}+\bH_{{\rm dip}}+\bH_{{\rm ex}}.$
$\bH_{{\rm ex}}=J\nabla^2\bM$
is the effective field due to direct exchange;
$\bH_{{\rm an}}=K \bM_0$
denotes the influence of anisotropy energy;
$\bH_{{\rm e}}$ represents the external field; and
$\bH_{{\rm dip}}$ denotes the
dipole-dipole interaction.
$\tau$ is the relaxation time, describing the relaxation of the
system towards its local equilibrium value of
magnetization.
Eq. (1) and (2) involves magnetization and charge currents,
to which we turn our attention.

To derive the transport equation for the vector
magnetization, we consider a local coordinate system
so that the electrons are quantized in the $y'$ direction.
We write down the transport equations in this local coordinate
system and
then rephrase it in covariant form, thus obtaining the general
transport equation.
According to Fick's law, the current for spin $s$ is given by
\begin{equation}
\bJ_s= \frac{n_s \mu_s}{e} \Bigl[ -e\nabla V + s\mu_B\nabla
H_{y'}\Bigr]-D_s\nabla n_s,
\end{equation}
where $\mu_B$ denotes the Bohr magneton,
$n_s$, $D_s$, and $\mu_s$ are the number density of charge carriers,
the diffusion coefficient, and the electron mobility, respectively.
$e<0$ is the electron charge, $s=\pm$ denote different spin
orientations, and $V=V_e+W$, with $V_e$ the electric potential
describing the external electric field and $W$ the local electric
(screening) potential due to the other electric charges determined
self-consistently by
\begin{equation}
W(\br)=\int d^3\br' U(\br-\br') \delta\! n(\br')
\end{equation}
with $U$ the Coulomb potential.
The total number density of charge carriers
and $y'$ component of magnetization are given by
$n=n_+ + n_-, M_{y'}=\mu_B (n_+ - n_-).$

From (3), we obtain expressions of the total charge current $\bJ_e$
and magnetization current $\tJ$. They are, after linearization and
rephrasing in covariant form, given by
\begin{equation}
\begin{array}{l}
\bJ_e=-\sigma\nabla V -\alpha_1 e \xi D\nabla \delta\! n
      -\displaystyle\frac{e}{\mu_B}\beta \xi D \nabla (\delta\!\bM\cdot \bp_0) \\
\tJ=- \sigma_M \nabla ( V \bp_0)
    - \xi D {\nabla\delta\!\bM}
    - \alpha_2\beta \mu_B \xi D \nabla ( \delta\! n  \bp_0)
\end{array}\label{JeJM}
\end{equation}
where
$\sigma=e\displaystyle\sum_s n_s\mu_s$,
$\sigma_M=\mu_B\displaystyle\sum_s s n_s\mu_s$,
$\beta=\displaystyle\frac{e\sigma_M}{\mu_B\sigma}$,
$\bp_0=\displaystyle\frac{\bM_0}{|\bM_0|}$ with
$\bM_0$ the local equilibrium magnetization, and
$\xi=\displaystyle\frac{\mu_B^2}{e^2}\displaystyle\frac{\sigma}{\chi D}$
with $D=\frac{1}{2}\displaystyle\sum_s D_s$
and $\chi$ the Pauli susceptibility at the Fermi energy.
$\alpha_1$ and $\alpha_2$ are two renormalized phenomenological
parameters, with $\alpha_1\approx\alpha_2\approx1$ in the limit
of strong ferromagnet.

Substituting the expression for $\tJ$ into the Landau-Lifshitz 
equation (2) we obtain the relaxation equation for $\bM$:
\begin{equation}
\nabla^2\delta\!\bM - \displaystyle\frac{1}{l_{sf}^2}\delta\!\bM
+\zeta\bp_0\times(\nabla^2\delta\!\bM-\frac{\kappa}{l_{sf}^2}\delta\!\bM)
= -\beta\mu_B  \bp_0 ( \alpha_3\nabla^2\delta\! n
                  -\frac{\delta\! n}{\lambda_0^2})
\label{eq1}
\end{equation}
where only $\bH_{{\rm ex}}=J\nabla^2\bM=J\nabla^2\delta\!\bM$ and
$\bH_{{\rm an}}=K\bM_0$ are
kept in the precession term $\gamma \bM\times \bH$, and use has been
made of Gauss' law:
$\nabla^2V=\nabla^2W=-\displaystyle\frac{e}{\epsilon_0}\delta\!n$.
The bare spin diffusion length $l_{sf}$ and the bare screening length
$\lambda_0$ are given by
$$
l^2_{sf}=\tau \xi D \ \ \mbox{\rm and} \ \
\lambda_0^2=\frac{\epsilon_0\xi D}{\sigma}
$$
respectively. Other dimensionless parameter are
$\zeta=\dfrac{\gamma|\bM_0| J}{\xi D}$ and
$\kappa=\dfrac{l_{sf}^2 K}{J}$, while
$\alpha_3$ is again a
renormalized phenomenological parameter collecting all
$\nabla^2 \delta\! n$ dependence including the change of $D$
(such as is caused by the dipole layer
at zero bias due to the difference of the work functions of the
material on the left and the right) near the interface.

The charge current conservation (1) yields, for the steady state,
\begin{equation}
\frac{\mu_B}{\lambda^2_0}\delta\! n
 - \alpha_1 \mu_B \nabla^2  \delta\! n
 -\beta \nabla^2 (\delta\! \bM \cdot \bp_0)=0
 \label{eq2},
\end{equation}
which, together with Eq.(\ref{eq1}),
describes the distribution of the charge and
magnetization on opposites sides away from the tunnel junction
in terms of their values at the junction. The values of the charge
and magnetization densities at the junction
can be determined by matching boundary conditions across the barrier.
We first solve these equations in the metal part of the junction. These
solutions determine the charge and magnetization dipole layers.

We expect the charge and magnetization dipole layers to decay away from the
interface with length scales controlled by the spin diffusion length and
the screening length.
Because of the vector nature of the magnetization, there are three normal
modes by which they can decay away from the interface.
Including the charge degree of freedom, there are four normal modes that
one can consider.
For ferromagnetic metal on the right hand side,
we thus consider the following ansatz:
\begin{equation}
\delta\! n^R=\sum_{i=1}^4 \delta\! n_{i0}^R
e^{-(z-\frac{d}{2})/l_{i}},\ \
\delta\!\bM^R=\sum_{i=1}^4 \delta\! \bM_{i0}^R
e^{-(z-\frac{d}{2})/l_{i}},   \label{ansatz}
\end{equation}
where the superscript $R$ denotes the right hand side.
Inserting these into (\ref{eq1}-\ref{eq2}) and letting the coefficients before
the exponential scaling functions vanish
for steady-state solutions, we get the renormalized decay lengths
\begin{equation}
l_1=\sqrt{\dfrac{\alpha_1-\alpha_3\beta^2}{1-\beta^2}}\,\lambda_0,\
l_{2}=\sqrt{1-\beta^2}\,l_{sf}, \
l_{3}=(a_+ + i a_-)l_{sf},\    l_4=l_3^*=(a_+ - i a_-)l_{sf}
\end{equation}
with
$$
a_{\pm}=\left[\dfrac{\sqrt{(1+\zeta^2\kappa^2)(1+\zeta^2)}\pm(1+\kappa\zeta^2)}
                     {2(1+\zeta^2\kappa^2)}\right]^{\frac{1}{2}}.
$$
As we indicated in the introduction, the screening length and
the spin diffusion length are renormalized.
As we shall see below, $l_3$ and $l_4$ correspond
the length scales with which the
``precession'' dies away from the interface.
The charge densities can be related to the magnetization densities by
\begin{equation}
\delta\! n_{10}^R=
\dfrac{1-\beta^2}{\beta(\alpha_1-\alpha_3)}\dfrac{\delta\!M_{10}^R}{\mu_B},\
\delta\! n_{20}^R=\dfrac{\lambda_0^2}{l_2^2}\dfrac{\beta\delta\!M_{20}^R}{\mu_B},\
\delta\! n_{30}^R=\delta\! n_{40}^R=0.
\label{n2M}
\end{equation}
Note that $\delta\! n_{20}^R$ is much smaller than $\delta\! M_{20}^R$ because
$\lambda_0<<l_2$.
Inserting the ``eigen-solutions'' into equations (\ref{ansatz}),
we finally obtain analytic expressions for the dipole layers:
\begin{eqnarray}
\delta\! n^R&=&
        \delta\! n^R_{10}e^{-(z-\frac{d}{2})/l_1}
       +\delta\! n^R_{20}e^{-(z-\frac{d}{2})/l_{2}}
                \\
\delta\!\bM^R&=&
               \bp_0^R \delta\! M^R_{10} e^{-(z-\frac{d}{2})/l_1}
       +\bp_0^R \delta\! M^R_{20} e^{-(z-\frac{d}{2})/l_{2}}
       +\bq_m^R \delta\! M^R_{30} e^{-\eta   (z-\frac{d}{2})/l_{sf}}
\end{eqnarray}
where
\begin{equation}
\eta  =\frac{a_+}{c^2},\   c^2=a_+^2+a_-^2, \
\bq_m^R=\sin\varphi'\sin\Phi \be_x -
        \sin\varphi'\cos\Phi \be_y +
        \cos\varphi' \be_z,\
\varphi'=\varphi^R + \frac{a_- z}{ c^2 l_{sf}}.
\end{equation}
$\delta\! M^R_{i0}$, with $i=1,2,3$,
and $\varphi^R$ are to be determined later.
Terms of the order $(\lambda_0/l_{sf})^2$
have been neglected since $l_{sf}^2>>\lambda_0^2$.
As advertised, the charge dipole layer is the sum of two terms, one decaying
with a length scale of the screening length; the other, the spin
diffusion length. The vector
magnetization dipole is now a sum of three terms. The first two (
$\delta\!\bM_{10} ^R$,  $\delta\!\bM_{20} ^R$ ) are
along the direction of the original magnetization; the last one is
perpendicular to the direction of the original magnetization
and describes the precession of the magnetization
around the original axis. Again, the first
two terms correspond to decay lengths of the order of the spin diffusion
length and the screening length, while the precession term only decays with a
length scale of the order of the spin diffusion length.

With $\delta\! n^L$,
$\delta\!\bM^L$, $\bJ^L_e$, and $\tJ^L$ obtained similarly for
the ferromagnet on the left hand side,
we can next focus on the boundary condition across the insulator.
The charge neutrality condition
\begin{equation}
\int_{\frac{d}{2}}^\infty  \delta\! n^R d z +
\int^{-\frac{d}{2}}_{-\infty}  \delta\! n^L d z =0
\end{equation}
yields
\begin{equation}
l_1  (\delta\! n^R_{10}+\delta\! n^L_{10}) +
l_{2}(\delta\! n^R_{20}+\delta\! n^L_{20})=0 \label{Q0}.
\end{equation}
We have, for simplicity,
assumed the identical parameters $D^R=D^L=D$,
$\xi^R=\xi^L=\xi$, $\sigma^R=\sigma^L=\sigma$, etc.
The continuity of charge current $\bJ_e$ and the spin current $\tJ$
implies
\begin{equation}
\bJ_e^L\biggl|_{z=-d/2}=\bJ_e^R\biggl|_{z= d/2}\equiv\bJ_e^0, \hskip 20pt
\tJ^L\biggl|_{z=-d/2}=\tJ^R\biggl|_{z= d/2}\equiv \tJ^0.
\end{equation}
The charge and magnetization currents (\ref{JeJM}) can be expressed
in terms of the $\delta\! M_{10}^R$, etc as:
\begin{eqnarray}
\bJ^R_e &=& \sigma \bE_{ext}   \label{Je0}          \\
\tJ^R   &=& \sigma \bE_{ext} \frac{\mu_B}{e} \beta   \bp_0^R +
       \frac{\beta \xi D\mu_B(1-\alpha_3)}{l_1}
        \be_z \bp_0^R \delta\! n^R_{10}
        e^{-(z-\frac{d}{2})/ l_1} \nonumber \\
       & &
      + \frac{(1- \beta^2)\xi D}{l_{2}} \be_z \bp_0^R \delta\! M^R_{20}
        e^{-(z-\frac{d}{2})/l_{2}}
      + \frac{\xi D}{c\, l_{sf}}\be_z
      \bq_j^R \delta\! M^R_{30} e^{-\eta   (z-\frac{d}{2})/l_{sf}} \label{JM0}
\end{eqnarray}
where $\bE_{ext}=E_{ext}\be_z$ is the external electric field inside
the conductor,
and
\begin{equation}
\bq_j^R=\sin\varphi''\sin\Phi \be_x -
        \sin\varphi''\cos\Phi \be_y +
        \cos\varphi'' \be_z,  \ \ \ \ \
\varphi''=\varphi'-\sin^{-1}\frac{a_-}{c}.
\end{equation}
Note that $\bq_m\cdot\bp_0=\bq_j\cdot\bp_0=0$.
The currents $\bJ_e$ and $\tJ$ are assumed to satisfy the boundary
conditions
\begin{eqnarray}
\be_z\cdot \bJ_e^0&=& 
      -\frac{1}{e r d}(G_S^R  \delta\! n^R -
                     G_S^L  \delta\! n^L )
       -\frac{1}{e r d\mu_B}(G_D^R  \delta\!\bM^R \cdot \bp_0^R -
                          G_D^L  \delta\!\bM^L \cdot \bp_0^L )
      \label{JeLR}                          \\
\be_z\cdot\tJ^0 &=&
  -\frac{\mu_B}{e^2 r d}(G_D^R \delta\! n^R\bp_0^R  - G_D^L \delta\! n^L
  \bp_0^L )
  -\frac{1}{e^2 r d}(G_S^R \delta\!\bM^R  - G_S^L \delta\!\bM^L )
   \label{JMLR}
\end{eqnarray}
where $G_D^R$ and $G_S^R$ ($G_D^L$ and $G_S^L$) are
phenomenological parameters describing combinations of inverse
densities of states of the ferromagnet on the right (left) hand
side \cite{Gset},
and $r$ is the resistivity of the junction (insulator).
The $\Delta W$ terms are much smaller and neglected for the
system.

>From equations (\ref{Q0}), (\ref{Je0}-\ref{JM0}), and (\ref{JeLR}-\ref{JMLR})
we can solve for $\delta\! n^L_{10}$, $\delta\! n^R_{10}$,
$\delta\! M^L_{20}$, $\delta\! M^R_{20}$,
$\delta\! M^L_{30}$, $\delta\! M^R_{30}$.
Since
$$
\dfrac{\l_1}{l_2}\sim \frac{\lambda_0}{l_{sf}}<<1, \ \ \
\dfrac{G_{S(D)}^{L(R)}}{D e^2 r}\sim\dfrac{1}{r\sigma}<<1,
$$
terms of higher order are neglected.
We find that the dominating term
$\delta\! M_{20}^L$ is proportional to the voltage $rdJ_e$.
The resistance of the junction, and not that of the metal
enters into consideration.  More specifically
\begin{equation}
\delta\! M_{20}^L=
-\delta\! M_{20}^R=
\dfrac{e \mu_B  r  d J_e}{G_D^L+G_D^R}. \\
\label{dM2}
\end{equation}
All other terms such as
$\delta\! M_{10}^L$,  $\delta\! M_{30}^L$, and $\delta\! n_{20}^L$
are much smaller than $\delta\! M_{20}^L$.
They are given by
\begin{equation}
\begin{array}{lllll}
\delta\! M_{10}^L=
-\delta\! M_{10}^R
=-\dfrac{l_1}{l_2}
                   \dfrac{\alpha_1-\alpha_3}{\alpha_2 -\alpha_3}
                   \delta\!M_{20}^L, &
\delta\! M_{30}^L=
 \dfrac{G_S^R}{G_S^L}\delta\! M_{30}^R
=\dfrac{c\,l_{sf} G_S^R \sin\Phi}{\xi D e^2 r d} \delta\!M_{20}^L, &
\varphi^L=
\varphi^R=\cos^{-1}\dfrac{-a_-}{c}
\end{array}
\end{equation}
where it has been assumed that
$\bp_0^L=\be_x$,
$\bp_0^R=\cos\Phi \be_x + \sin\Phi\be_y$.
The charge densities are related to the magnetization densities through
Eq.(\ref{n2M}).

Eq.(\ref{dM2}) is our main result.
It shows that an effective switching field $H_s
=\delta M^R_{20}/\chi$ can be generated
which is basically controlled by the
voltage and, when normalized with respect to current, is much bigger
than that caused by spin accumulation picture.
By controlling the sign of $J_e$, the sign
of this field can be changed. In this calculation, we have assumed
the magnetization is uniform in the $x$-$y$ plane. In a small element,
there are edge domains where the switching first starts.
We expect that this effective field will also act on the edge domains.

There is some experimental evidence of a much stronger current
induced switching effect than expected based on the spin
accumulation picture. Kohlstedt and coworkers \cite{Kohl} have
recently studied experimentally a three terminal multiple junction
$F_1-I_1-F_2-I_2-F_3$ with leads at $F_1$, $F_2$ and $F_3$. They
found that the magnetoresistance for $F_2-I_2-F_3$ is changed as a
current is passed through $F_1-I_1-F_2$. They interpret the change
in the magnetoresistance ratio in the second tunnel junction as a
change  in the magnetization in $F_2$ when a current is passed
through the first tunnel junction. They found that the change in
magnetization in $F_2$ is {\bf much bigger than that caused by the
spin accumulation effect}! Furthermore, the change in the
resistance of the second junction is {\bf increased} as the
resistance of the first junction is increased, {\bf opposite} to
the trend according to the spin accumulation picture. Their result
is consistent with the present picture.

In conclusion, We have studied the transport of the vector
magnetization through a magnetic tunnel junction. Because of the
induced charge and magnetization dipole layers at the interfaces,
we found in tunnel structures the possibility of an enhanced spin
switching effect that, when normalized with respect to the current, is
much bigger than that caused by the spin accumulation picture.

Z L was supported in part by the Chinese NSF.


\begin{references}
\bibitem{Gset}
 In a simple model, they are given by:
$G_S^{L,R}=\frac{1}{2N_S^{L,R}}$ and
$G_D^{L,R}=\frac{1}{2N_D^{L,R}}$, with
$\frac{1}{N_S}=\frac{1}{N_+}+\frac{1}{N_-}$ and
$\frac{1}{N_D}=\frac{1}{N_+}-\frac{1}{N_-}$,
where $N_+$ and $N_-$ are density of states
at the Fermi energy of spin up and down, respectively.
\end{references}
\end{document}